\documentclass[a4,reqno,10pt]{amsart}

 \usepackage{amsmath,amsthm,amssymb,xcolor,graphicx,soul}

\usepackage{amscd,verbatim,a4wide}
\usepackage[all]{xy}

\usepackage{hyperref}

\def\L{{\mathcal L}}

\def\jp{\frac{1}{2}}
\def\be{\begin{equation}}
\def\ee{\end{equation}}
\def\wedgec{\stackrel{\wedge}{,}}

\theoremstyle{plain}

\theoremstyle{definition}

\theoremstyle{remark}

\numberwithin{equation}{section}
\numberwithin{theorem}{section}
\numberwithin{figure}{section}
\numberwithin{table}{section}

\setstcolor{red}

\usepackage{tikz}
\usetikzlibrary{arrows}
\usetikzlibrary{positioning}
\usepackage{amsmath}
\usepackage{amsfonts}
\usepackage{hyperref}
\usepackage{amsthm}

\newcommand{\g}{\mathfrak{g}}
\newcommand{\GG}{\mathsf{G}}

\newcommand{\HH}{\mathsf{H}}

\newcommand{\R}{\mathbb{R}}
\newcommand{\pr}{\partial}

\begin{document}

\title[Hidden isometry of  ``T-duality without isometry"]{Hidden isometry of  ``T-duality without isometry"}

\author[P Bouwknegt]{Peter Bouwknegt}
\address[P Bouwknegt]{
Mathematical Sciences Institute,
Australian National University, 
Canberra, ACT 2601, Australia}
\email{peter.bouwknegt@anu.edu.au}

\author[M Bugden]{Mark Bugden}
\address[M Bugden]{
Mathematical Sciences Institute,
Australian National University, 
Canberra, ACT 2601, Australia}
\email{mark.bugden@anu.edu.au}

\author[C Klim\v c\'\i k]{Ctirad Klim\v c\'\i k}
\address[C Klim\v c\'\i k]{
Institut de Math\'ematiques de Luminy,
Aix Marseille Universit\'e, CNRS, Centrale Marseille I2M, UMR 7373,
13453 Marseille, France}
\email{ctirad.klimcik@univ-amu.fr}

\author[K Wright]{Kyle Wright}
\address[K Wright]{
Department of Theoretical Physics,
Research School of Physics and Engineering, and
Mathematical Sciences Institute,
Australian National University, 
Canberra, ACT 2601, Australia}
\email{wright.kyle.j@gmail.com}

\begin{abstract}
We  study  the T-dualisability criteria  of  Chatzistavrakidis, Deser and Jonke \cite{CDJ15}  who recently
used Lie algebroid  gauge theories to obtain sigma models exhibiting 
a ``T-duality without isometry".  We point out that those  T-dualisability criteria 
are not written invariantly in  \cite{CDJ15} and depend on the choice of the 
algebroid framing. We then show that there  always exists an isometric  framing
for which the  Lie algebroid gauging boils down to  standard Yang-Mills gauging. The
``T-duality without isometry" of Chatzistavrakidis, Deser and 
Jonke is therefore nothing but traditional  isometric  non-Abelian T-duality in disguise.
\end{abstract}

\maketitle


\section{Introduction}

T-dualisability is a rare property of non-linear sigma-models and it is not known what necessary conditions must be imposed on a target space metric 
$G$, and closed 3-form field $H$, such that the corresponding sigma-model has a T-dual with $(\widehat G,\widehat H)$. 
On the other hand,  several sufficient conditions are known, giving various T-dualities like the Abelian one \cite{KY,SaS} or non-Abelian 
one \cite{FJ,FT,DQ,AAL94}, both in turn included as special cases of Poisson-Lie T-duality \cite{KS95,Kli95}.   Chatzistavrakidis, Deser and 
Jonke (CDJ in what follows) recently proposed a new set of sufficient conditions which, they claimed,  would give rise to new examples of T-dual pairs \cite{CDJ15}.  
Their dualisability conditions appear much less restrictive than those previously described in T-duality research. It is the purpose 
of the present work  to show that, in reality,  they are not less restrictive as  they  give rise to the same duality pattern as that of traditional 
non-Abelian T-duality.

The  proposal of CDJ  for dualising a given sigma model on a target $M$, is an extension of  the  
Ro\v cek-Verlinde approach \cite{RV,DQ},  which amounts to the introduction of an intermediate gauge theory yielding  the 
T-dual pair of sigma models upon eliminating different sets of fields.  It was traditionally thought that  the Ro\v cek-Verlinde 
intermediate gauge theory can be constructed only if the background  of the sigma-model is isometric with respect to the  
action of the Lie algebra $\mathfrak g$  of the gauge group. However, CDJ have argued that 
more general gaugings are possible if one uses 
 the recently introduced Lie algebroid gauge theory \cite{Str04, MS09,KS14}.   The construction of the Lie algebroid generalisation of the 
 Ro\v cek-Verlinde intermediate gauge theory  requires  the existence of a Lie  algebroid bundle $Q$, over the target $M$, as well as  a 
 fixed connection $\nabla^\omega$  on $Q$ compatible with the sigma model background. As CDJ  show,  
 the compatibility of $\nabla^\omega$, $G$ and $H$   can be expressed in a particularly simple way  for exact 3-form backgrounds 
 $H=dB$ where it reads:
\be 
\L_{\rho(e_a)}G=\omega^b{}_a\vee\iota_{\rho(e_b)}G,\quad \L_{\rho(e_a)}B=\omega^b{}_a\wedge \iota_{\rho(e_b)}B\,.   \label{1}
\ee
Here $e_a$ form local frames of the Lie algebroid, the Lie derivatives are taken with respect to the
anchored frames $\rho(e_a)$, the symbols $\vee$ and $\wedge$ stand respectively for the symmetrised and 
anti-symmetrised direct products of 1-forms on $M$, and the 1-forms $\omega^b{}_a$ are defined by the relations
\be
\nabla^\omega e_a:=\omega^b{}_a\otimes e_b\,.
\ee
Since the choice of the connection $\nabla^\omega$ seems largely arbitrary, it may appear from (\ref{1}) that 
a vast set of non-isometric backgrounds could be gauged, thus producing a new and rich T-duality pattern. 
However, as we shall argue in this paper, this is not the case. The simplest way to understand what is happening is to realise 
that the compatibility conditions (\ref{1}), as given by CDJ
in Ref.\ \cite{CDJ15} are not written invariantly; upon a local changes of frames $e'_a=P^b{}_a e_b$, $P^b{}_a \in C^\infty(M)$, they change to 
\be 
\L_{\rho( e'_a)}G=  \omega'^{b}{}_a\vee\iota_{\rho( e'_b)}G,\quad \L_{\rho(e'_a)}B=\omega'^{b}{}_a\wedge \iota_{\rho(e'_b)}B\,,  \label{1a}
\ee
where 
\be 
\nabla^{\omega'} e'_a := \omega'^b{}_a\otimes e'_b \,.
\ee
The components of the connection form $\omega^b{}_a$ transform non-homogeneously upon a change of the framing, and we may naturally 
question whether there exists a distinguished frame $\hat e_a$ for which  they all vanish. This question can be answered in the affirmative, and this fact 
follows from  the Lie algebroid gauge invariance of the Ro\v cek-Verlinde intermediate gauge theory.
It is therefore always possible  to write down an equivalent version  of  the CDJ   compatibility conditions (\ref{1})  
in the standard isometric form
\be 
\L_{\rho(\hat e_a)}G=0\,,\quad \L_{\rho(\hat e_a)} B = 0   \,.\label{2}
\ee
Moreover, the gauge invariance of the intermediate gauge theory also requires that the structure functions
$\hat C^c{}_{ab}$ defined by the Lie algebroid brackets 
\be
[\hat e_a,\hat e_b] \equiv \widehat C^c{}_{ab}\  \hat e_c ,
\ee 
be constants, and we thus recover the standard intermediate Yang-Mills gauge theory leading 
to traditional non-Abelian T-duality \cite{FJ,DQ}.

The plan of our paper is as follows: in Section 2 we expose some useful preliminary background on traditional non-Abelian T-duality.
In Section 3 we review 
the ``T-duality without isometry" proposal of CDJ and detail the field redefinitions which reproduce standard non-Abelian T-duality.    
In Section 4 we work out  the case of non-exact $3$-form background $H$. In Section 5
we provide a geometric interpretation of the field redefinitions  from the invariant perspective of Lie algebroid gauge theory. In Section 6, we 
{illustrate a few examples where, by simple field redefinitions,} the traditional isometric Ro\v cek-Verlinde gauge theory  may look  like  a 
non-trivial Lie algebroid gauge theory. In particular, we unmask the ``non-isometric T-duality" example
of CDJ   presented in \cite{CDJ15}.
Finally, we end with a short discussion.


\section{Preliminaries on the non-Abelian T-duality}

To set up some technical and notational background, as well as remind the reader of the gauging approach  approach to 
T-duality, we review traditional non-Abelian T-duality   obtained by the Ro\v cek-Verlinde procedure \cite{FJ,FT,DQ}.  
We  first restrict our attention to backgrounds for which $H=dB$ is an exact  $3$-form, postponing the study of  
cohomologically non-trivial backgrounds to Section 4. 

Let a Lie group $\mathsf G$ act from the right on the target manifold $M$, let $T_a$ be a basis
of the Lie algebra
$\mathfrak g\equiv \text{Lie}(\mathsf G)$,  and
$v_a$ the set of vector fields on $M$  corresponding to  the infinitesimal right  actions of the elements $T_a$.   
The Lie derivatives $\L_{v_a}v_b$ then satisfy
\be 
\L_{v_a}v_b = [v_a,v_b] = C^c{}_{ab} \, v_c\,, \label{sc} 
\ee
where $C^c{}_{ab}$  are the  structure constants  of $\mathfrak g$  in the basis $T_a$.

Denoting the (Lorentzian) cylindrical world-sheet by  $\Sigma$
and introducing coordinates $X^i$ on $M$, 
we write the sigma model action with the background metric $G$ and the 3-form field $H=dB$ as
\be 
S(X^i)= \jp \int_{\Sigma}   dX^i \wedge \left(G_{ij}*dX^j+B_{ij}  dX^j\right) \,.    \label{4}
\ee 
Here 
 $d$ denotes the de Rham differential,  $*$ ($*^2=1$) the Hodge star on the world-sheet $\Sigma$, and the $X^i$ are 
viewed as functions on $\Sigma$ describing a string moving in $M$.

If the Lie derivatives of the metric and the $B$ field vanish
\be 
\L_{v_a}G=0,\quad \L_{v_a}B=0   \,, \label{2'}
\ee 
then the sigma model (\ref{4}) can be gauged in the standard Yang-Mills way. This means
that one introduces a  world-sheet one-form $A$ valued in the Lie algebra $\mathfrak g\equiv$Lie($\mathsf G$), 
  a world-sheet scalar $\eta$ valued in the dual $\mathfrak g^*$, and the gauged action
\be 
S(X^i,A,\eta)= \jp \int_{\Sigma}   DX^i \wedge \left(G_{ij}*DX^j+B_{ij}  DX^j\right)
+ \int _\Sigma \langle \eta,F(A) \rangle \,. \label{5}
\ee
Here $\langle \cdot\ ,\cdot \rangle$ is the canonical pairing between $\mathfrak g^*$ and $\mathfrak g$, $F(A)$ is the
standard Yang-Mills field strength
\be 
F(A) := dA+A\wedge A \equiv \left(dA^a+\jp C^a{}_{bc}\,A^b\wedge A^c\right)T_a\,,
\ee
and $DX^i$ are the covariant derivatives
\be 
DX^i:=dX^i-v_a^iA^a \,.   \label{cd}
\ee
If the isometry conditions (\ref{2'}) hold, the action (\ref{5}) is gauge invariant with respect to the following local infinitesimal gauge transformations:
\be 
\delta_\epsilon X^i=v^i_a\epsilon^a, \quad 
\delta_\epsilon A =d\epsilon +[A,\epsilon]\equiv\left(d\epsilon^a+C^a{}_{bc}A^b\epsilon^c\right)T_a,\quad 
\delta_\epsilon \eta=-{\rm ad}^*_\epsilon\eta\equiv -C^c{}_{ab}\eta_c \epsilon^bT^{*a}  \,.
\ee
Here $\epsilon$ is a function on the world-sheet valued in $\mathfrak g$, and ${\rm ad}^*$ denotes the co-adjoint action of $\mathfrak g$ on $\mathfrak g^*$.
   
Varying the Lagrange multiplier $\eta$  {forces the field strength to vanish, thereby imposing that}
the gauge field $A$ be pure gauge $A=-dg\, g^{-1}$, and 
the   action (\ref{5}) becomes that of the original model (\ref{4})
\be
S(X^i,-dg\ g^{-1},\eta)= \jp \int_{\Sigma}  dY^i \wedge \left(G_{ij}*dY^j+B_{ij}  dY^j\right) \,.
\ee
Here $Y^i=\ ^gX^i$ which means that $Y^i$ is obtained from $X^i$ by applying  the gauge transformation $g$. If instead, we eliminate  
the non-dynamical fields $A$ from (\ref{5}), as well as fixing the gauge, we obtain  the dual sigma model. The exact form of the dual action depends on several 
factors; like whether or not the $\mathsf G$ action on $M$ is free, or the presence of so-called spectator fields.  
We  do not give the complete account of all possible cases (the interested reader can find it in \cite{AAL94,AABL,DQ,GR}),   because our 
concern is different. We will show that the Lie algebroid generalisation of the intermediate gauge theory proposed by 
CDJ in \cite{CDJ15} can be rewritten,  using appropriate field redefinitions, in the standard 
non-Abelian T-duality  form (\ref{5}).  It follows that the CDJ proposal cannot describe more general T-duality patterns than that of traditional non-Abelian T-duality.
    

 \section{CDJ gauge theory}
 
CDJ  generalised the structural data $M,G,B,v_a$ considered in  Subsection 2.1  
by including an additional  matrix valued $1$-form    $\omega^a{}_{b}\equiv \omega^a{}_{bi}dX^i$ on $M$, and by promoting the 
structure constants $C^c{}_{ab}$ to functions on $M$.
The action of the  intermediate gauge theory is then proposed to be  the following expression:\footnote{The formulas appearing
here are for exact 3-form $H=dB$, and are equivalent to the equations appearing in \cite{CDJ15} under this assumption. 
Non-trivial $H$, as considered in \cite{CDJ15}, is treated in Section 4. }
\be 
S(X^i,A,\eta)= \jp \int_{\Sigma}   DX^i \wedge \left(G_{ij}*DX^j+B_{ij}  DX^j\right)
+ \int _\Sigma \langle \eta,F_\omega(A,X)\rangle\,, \label{2.1}
\ee
where the covariant derivatives $DX^i$ are as before (cf.\ (\ref{cd})) and the generalized field strength $F_\omega(A,X)$   
(borrowed from Ref. \cite{KS14})   is given by the formula
\be 
F^a_\omega(A,X):=dA^a+\jp C^a{}_{bc}(X)A^b\wedge A^c-\omega^a{}_{bi}A^b\wedge (dX^i-v_c^iA^c)\,.   \label{cc}
\ee
CDJ  then argued that a necessary condition for the infinitesimal  gauge invariance of the theory (\ref{2.1}), 
required for T-duality applications, is given by 
\be 
\L_{v_a}G=\omega^b{}_a \vee\iota_{v_b}G\,, \quad \L_{v_a}B = \omega^b{}_a  \wedge \iota_{v_b}B  \,.    \label{gic} 
\ee 
The  infinitesimal gauge transformations themselves depend on  $\omega^a{}_{b}$ and they read (cf.~\cite{CDJ15})
\begin{align}
\delta _\epsilon X^i &  = v_a^i\epsilon^a\,,  \nonumber \\
\delta _\epsilon A^a & = d\epsilon^a+C^a{}_{bc}\,   A^b\epsilon^c+\omega^a{}_{bi}(dX^i-v^i_aA^a)\epsilon^b\,,  \label{gt2} \\
\delta _\epsilon \eta_a & = \left(-C^c{}_{ab} \eta_c+v^i_a\omega^c{}_{bi}\eta_c\right) \epsilon^b\,.  \nonumber
\end{align}
We point out that the conditions (\ref{gic}) are not sufficient to guarantee the gauge invariance. 
This can be seen by evaluating the variation $\delta _\epsilon F_\omega(A,X)$ of the field strength:
\be 
\delta _\epsilon F^a_\omega(A,X)=(d\omega^a{}_b+\omega^a{}_c\wedge\omega^c{}_b)\epsilon^b+O(A) +O(A^2)\,,
\ee
where $O(A)$ and  $O(A^2)$ stand for the terms linear and quadratic in $A$, respectively.  The variation $\delta _\epsilon \langle \eta,F_\omega(A,X)\rangle$ is required to vanish,
$$ 
0=\delta _\epsilon \langle\eta,F_\omega(A,X)\rangle = \langle \delta _\epsilon \eta,F_\omega(A,X) \rangle  + \langle \eta,\delta _\epsilon  F_\omega(A,X) \rangle 
$$
\be
=\eta_a(d\omega^a{}_b+\omega^a{}_c\wedge\omega^c{}_b)\epsilon^b+O(A) +O(A^2)\,.
\ee
All three terms
must vanish separately which means that  the conditions (\ref{gic}) of the gauge 
invariance have to be supplemented by, at least, one other one
\be 
d\omega^a{}_b+\omega^a{}_c\wedge\omega^c{}_b=0\,.\label{mc}
\ee
The condition (\ref{mc}) is easy to solve since it has  the Maurer-Cartan form. Therefore, there must exist a matrix $K^a{}_b(X)$ such that 
\be 
\omega^a{}_b=(K^{-1})^a{}_c \, dK^c{}_b \,.\label{zc}
\ee
It turns out  that the conditions (\ref{gic}),  together with (\ref{mc}), are necessary but still not sufficient to guarantee the 
gauge invariance. In order to find the full set of conditions to be imposed, 
we perform the following field redefinitions:
\be 
\widehat A^a=K^a{}_b A^b\,,\quad \hat\eta_a=\eta_b(K^{-1})^b{}_a \,.\label{fir}
\ee
In terms of the  new fields $X^i, \widehat A^a$ and $\hat\eta_a$, the action (\ref{2.1}) of  CDJ  acquires the following form:
\be 
S(X^i,\widehat A,\hat\eta)= \jp \int_{\Sigma}   DX^i \wedge \left(G_{ij}*DX^j+B_{ij}  DX^j\right)
 + \int _\Sigma \hat\eta_a (d\widehat A^a +\jp\widehat C^a{}_{bc}(X)\widehat A^b\wedge \widehat A^c)\,, \label{fr}
\ee
where
\be 
\widehat C^a{}_{bc} := K^a{}_d\left((K^{-1})^e{}_b(K^{-1})^f{}_c  C^d{}_{ef}+(K^{-1})^e{}_bv^i_e\partial_i(K^{-1})^d{}_c-(K^{-1})^e{}_cv^i_e\partial_i(K^{-1})^d{}_b\right)
\ee
and
\be 
DX^i = dX^i-v_a^iA^a=dX^i-\hat v_a^i\widehat A^a, \quad \hat v_a^i:= v^i_b (K^{-1})^b{}_a\,.   \label{cd'}
\ee
Furthermore, upon the field redefinitions $(A,\eta)\to(\widehat A,\hat\eta)$,  the gauge transformation formulas (\ref{gt2}) 
simplify
\begin{align}
\delta _{\hat\epsilon} X^i & =\hat v_a^i\hat\epsilon^a \,, \nonumber \\
\delta _{\hat\epsilon}\widehat A^a & = d\hat\epsilon^a+\widehat C^a{}_{bc}(X)\widehat A^b\hat\epsilon^c \,,  \label{gt'}   \\
\delta _{\hat\epsilon}\hat\eta_a & = -\widehat C^c{}_{ab}(X) \hat\eta_c\hat\epsilon^b\,,\nonumber
\end{align}
where 
\be 
\hat\epsilon^a=K^a{}_b\, \epsilon^b \,.\label{mar}
\ee
   
Remarkably, upon the field redefinitions,  the gauge invariance conditions (\ref{gic}) guaranteeing the gauge invariance 
of the first term in the action (\ref{fr}) become the isometry conditions 
\be 
\L_{\hat v_a}G=0,\quad \L_{\hat v_a}B=0,\label{2.2}
\ee
which can be established  directly from  (\ref{gic}):
\begin{equation*}
K^a{}_b \, \L_{\hat v_a}G=K^a{}_b \L_{v_c(K^{-1})^c{}_a}G=\L_{v_b}G-(K^{-1})^a{}_c \, dK^c{}_b\vee\iota_{v_a} G =
\L_{v_b}G-\omega^a{}_b\vee\iota_{v_a} G = 0 \,,
\end{equation*}
and similarly,
\begin{equation*}
K^a{}_b\, \L_{\hat v_a}B = K^a{}_b \L_{v_c(K^{-1})^c{}_a}B=\L_{v_b}B-(K^{-1})^a{}_c dK^c{}_b\wedge\iota_{v_a} B =
\L_{v_b}B-\omega^a{}_b\wedge\iota_{v_a}B = 0 \,.
\end{equation*}
Now we turn to the second term in the action (\ref{fr}), i.e. the one containing the Lagrange multiplier $\eta$.  It is  easy to calculate the variation
of the action (\ref{fr}) with respect to the gauge transformations
(\ref{gt'}) provided that the isometry conditions (\ref{2.2}) are fulfilled:
\be  
\delta _{\hat\epsilon}S(X^i,\widehat A,\hat\eta)=\int_\Sigma \hat\eta_a (\partial_i  \widehat C^a{}_{bc})\hat\epsilon^cDX^i\wedge \widehat A^b \,.
\ee
The gauge invariance thus require that the structure functions $\widehat C^a{}_{bc}$ be constants.\footnote{It is not difficult 
to conclude that there exists no modification of
the gauge transformation formula $\delta _\epsilon \eta_a= \left(-C^c{}_{ab}{}\eta_c+v^i_a\omega^c{}_{bi}\eta_c\right)\epsilon^b$ 
such that the modified gauge invariance criteria would permit any possibility other 
than  $ \omega^a{}_b=(K^{-1})^a{}_cdK^c{}_b$ and $\widehat C^a{}_{bc}$ constants.} 
We  observe that our  field redefinitions (\ref{fir}) permit us to rewrite the action (\ref{2.1}) of  CDJ gauge theory  precisely in the 
Ro\v cek-Verlinde form (\ref{5}) corresponding to isometric non-Abelian T-duality. Consequently, 
the approach of CDJ  cannot give any T-duality
pattern not already contained in the traditional non-Abelian T-duality story.


\section{Inclusion of non-exact $3$-form background $H$}

Let $H$ be a closed $3$-form on the target manifold $M$ which is not exact. In this case $H$ cannot  be written globally as $dB$ for any  
$2$-form field $B$  but we can introduce an auxiliary $2$-form field $C$ on $M$  and write down the following 
action:\footnote{Note that if $H=dB$ then $C$ can be identified with $B$ and the action
(\ref{4.1}) reduces to  (\ref{2.1}) as it should.}
\begin{align}
S(X^i,A,\eta) = & \jp \int_{\Sigma}   DX^i \wedge \left(G_{ij}*DX^j+C_{ij}  DX^j\right) 
 + \int_{\Sigma}   \langle \eta,F_\omega(A,X) \rangle \nonumber \\
& +\frac{1}{6} \int_{\Sigma_3}H_{ijk}dX^i\wedge dX^j\wedge dX^k-\jp\int_{\Sigma}C_{ij}dX^i\wedge dX^j  \,.\label{4.1}
\end{align}
Here $\Sigma_3$ is a volume for which the world-sheet $\Sigma$ is the boundary.

The action (\ref{4.1}) was introduced by CDJ in order to study
``T-duality without isometry" in the presence of the non-exact $3$-form background $H$.
The  gauge invariance conditions\footnote{The gauge invariance conditions originally written in  \cite{CDJ15},  or 
in \cite{CDJS16a}, use the notation $\theta_a:=-\iota_{v_a}C$ and  therefore look slightly different.}  
of the action  (\ref{4.1}) with respect to the gauge transformations (\ref{gt2}) read 
\be 
\L_{v_a}G=\omega^b{}_a\vee\iota_{v_b}G,\quad \L_{v_a}C=\omega^b{}_a\wedge \iota_{v_b}C\,, 
\quad \iota_{v_a}(H-dC)=0, \quad  d\omega^a{}_b+\omega^a{}_c\wedge\omega^c{}_b=0 \,.\label{gic'} 
\ee
The last condition can be solved as in Section 3, yielding
\be 
\omega^a{}_b = (K^{-1})^a{}_c\, dK^c{}_b\,.\label{zc'} 
\ee
With the field redefinitions 
\be 
\widehat A^a = K^a{}_b\, A^b,\quad \hat\eta_a=\eta_b(K^{-1})^b{}_a \,,\label{fir'}
\ee
the action  (\ref{4.1}) and the gauge transformations (\ref{gt2})  acquire the following form
\begin{align}
S(X^i,\widehat A,\hat\eta) = &  \jp \int_{\Sigma}   DX^i \wedge \left(G_{ij}*DX^j+C_{ij}  DX^j\right) 
 + \int _\Sigma \hat\eta_a (d\widehat A^a +\jp\widehat C^a{}_{bc}(X)\widehat A^b\wedge \widehat A^c) \nonumber \\
& +\frac{1}{6}\int_{\Sigma_3}H_{ijk}dX^i\wedge dX^j\wedge dX^k-\jp\int_{\Sigma}C_{ij}dX^i\wedge dX^j  \,,  \label{4.1'} 
\end{align}
and 
\begin{align}
\delta _{\hat\epsilon} X^i & =\hat v_a^i\hat\epsilon^a\,, \nonumber \\
\delta _{\hat\epsilon}\widehat A^a & = d\hat\epsilon^a+\widehat C^a{}_{bc}(X)\widehat A^b\hat\epsilon^c \,, \label{gtt'}  \\
\delta _{\hat\epsilon}\hat\eta_a & =  -\widehat C^c{}_{ab}(X) \hat\eta_c\hat\epsilon^b\,.\nonumber   
\end{align}
Here
\be
\hat\epsilon^a=K^a{}_b\, \epsilon^b \,, 
\ee
\be 
\widehat C^a{}_{bc}\equiv K^a{}_d\left((K^{-1})^e{}_b(K^{-1})^f{}_cC^d{}_{ef}
+(K^{-1})^e{}_bv^i_e\partial_i(K^{-1})^d{}_c-(K^{-1})^e{}_cv^i_e\partial_i(K^{-1})^d{}_b\right)  \,,
\ee
and
\be 
DX^i = dX^i-v_a^iA^a = dX^i-\hat v_a^i\widehat A^a\,, \quad \hat v_a^i =  v^i_b (K^{-1})^b{}_a\,.   \label{cd''}
\ee
As in Section 3,  the gauge invariance requires that the structure functions $\widehat C^a{}_{bc}$ be constants
and the conditions (\ref{gic'}) become the standard isometry conditions
\be 
\L_{\hat v_a}G=0\,, \quad \L_{\hat v_a}C=0\,, \quad \iota_{\hat v_a}(H-dC)=0  \,.
\ee
The gauge theory (\ref{4.1'}) thus boils down to the  Ro\v cek-Verlinde Yang-Mills theory underlying standard 
non-Abelian T-duality in the presence of the WZ term. 
   
 
\section{Lie algebroid gauged sigma models}
  
So far we have been using  the coordinates $X^i$ on $M$, in order to compare more directly
our calculations with those of CDJ. However, we have obtained our  principal insights invariantly, 
using the language of the Lie algebroid gauged sigma models \cite{KS14}. We consider it appropriate to  include this invariant 
perspective in the present paper, as it highlights the
geometric origin of the field redefinitions (\ref{fir}).
  
Recall that a Lie algebroid  is a vector bundle $Q$ over the manifold $M$, equipped with a Lie bracket $[\cdot\, ,\cdot]_Q$ 
on the space of sections $\Gamma(Q)$ and with an anchor homomorphism $\rho: Q\to TM$   intertwining the bracket
$[\cdot \,, \cdot]_Q$ with the standard Lie bracket of vector fields $[\cdot \, , \cdot]_{TM}$. Given a background metric $G$, a  $2$-form
$B$ on $M$,  and a connection $\nabla^\omega: \Gamma(Q)\to \Omega^1(M)\otimes \Gamma(Q)$ on the vector bundle  $Q$, we can  define
the  CDJ gauge theory  invariantly.   It  is a classical field theory  on the world-sheet $\Sigma$ 
with the action   
\be 
S(X,A,\eta)=\jp\int_\Sigma \vert\!\vert TX-\rho(A)\vert\!\vert_G^2+\int_\Sigma  X^*_AB+ 
\int_\Sigma \langle \eta, (X^*d^{\nabla^\omega})A-\jp \ ^Q{\mathcal T}^\omega(A\wedgec A)\rangle\,.\label{5.1}
\ee  
The dynamical fields  of the theory are: a map  $X:\Sigma\to M$,  a   $1$-form $A$ on $\Sigma$ 
with values in the pull-back bundle $X^*Q$  and a 
section  $\eta$ of the dual pull-back  bundle $X^*Q^*$.  The  expression 
$F_\omega (A,X):=(X^*d^{\nabla^\omega})A-\jp \ ^Q{\mathcal T}^\omega(A\wedgec A)$ takes values in $\Lambda^2T^*\Sigma\otimes X^*Q$ and 
it is conveniently referred to as the Lie algebroid field strength.    
  
Let us  explain in more detail the  notation.  First we note that the connection $\nabla^\omega: \Gamma(Q)\to \Omega^1(M)\otimes \Gamma(Q)$ 
on the vector bundle  $Q$ induces  the so-called linear connection $^Q\nabla^\omega: \Gamma(Q) \to \Gamma(Q^*) \otimes \Gamma(Q)$ 
on the Lie algebroid $Q$, defined by the relation
$$\ ^Q\nabla^\omega_{s_1}s_2:=\nabla^\omega_{\rho(s_1)}s_2, \quad s_1,s_2\in \Gamma(Q).$$
To the linear connection $^Q\nabla^\omega$ on $Q$ is then associated the $Q$-torsion   $^Q{\mathcal T}^\omega$  which  is $C^\infty(M)$-bilinear form on 
$\Gamma(Q)\times \Gamma(Q)$ with values in $\Gamma(Q)$:
\be \ ^Q{\mathcal T}^\omega(s_1,s_2):= \ ^Q\nabla^\omega_{s_1}s_2-\ ^Q\nabla^\omega_{s_2}s_1- [s_1,s_2]_Q.\ee
It must be mentioned  that by writing $^Q{\mathcal T}^\omega(A\wedgec A)$ as in Eqn.~(\ref{5.1}) 
we have used somewhat short-hand notation, with the purpose not to make the expression too heavy from
the notational point of view. In reality, we should have written rather
$X^* \ ^Q{\mathcal T}^\omega(\check{A}\wedgec \check{A})\big\vert_{X(u)}$, where $u\in\Sigma$ and $\check{A}$  
is any section of $\Omega^1(M,Q)$ with the property $X^*\check{A} \big\vert_{X(u)}=A(u)$ (here $X^*$ stands for the pull-back of differential forms and
$.\big\vert_{X(u)}$ means the restriction to the algebroid fiber over the point $X(u)$). It is  the crucial property of 
$C^\infty(M)$-bilinearity  of the torsion which guarantees that the ambiguity of the choice of the lifted section $\check{A}$ does not influence
the value of the field strength $F_\omega (A,X)$.
 
Recall  also that $X^*d^{\nabla^\omega}$ stands for  the standard pull-back of the extension of the connection 
$\nabla^\omega$  to the differential forms valued in $Q$ and $\vert\!\vert TX-\rho(A)\vert\!\vert_G$ means taking simultaneously the $G$-norm of the covariant
tangent map $TX-\rho(A)$ in $X^*TM$ and the Minkowski  (indefinite) norm of $1$-forms on the world-sheet 
$\Sigma$.  Finally, $X_A^*B$ stands for the covariant pull-back of the differential form, which is the $2$-form on 
$\Sigma$ given at every point $u\in \Sigma$ by contracting   $B$ in $X(u)$ 
with   $(TX-\rho(A))\wedge (TX-\rho(A))$.

To achieve our invariant description of CDJ theory,  we have to  define infinitesimal gauge transformations of the fields 
$X,A$ and $\eta$. The infinitesimal parameters $\epsilon$ of  this transformations must 
 be sections of  the pull-back bundle $X^*Q$ and the first of the transformations (\ref{gt2}) evidently reads
\be \delta_\epsilon X =  \rho(\epsilon).\label{igt1}\ee
To write invariantly the second and the third of the transformations (\ref{gt2}) is, however, more subtle, since putative invariant variations   
$\delta_\epsilon A$ or $\delta_\epsilon\eta$ do not make sense, because both $A$ and $\eta$ live in the pull-back bundles
which themselves change with the variation of $X$.  This means that, at a given $u\in\Sigma$, we cannot subtract the field $A(u)$ 
from the  transformed field $A_\epsilon(u)$ in order to define a variation $\delta_\epsilon A(u)$, being  unable to subtract
two vectors living in different spaces: $A(u)$ lives in the algebroid   fiber over $X(u)$ whereas  $A_\epsilon(u)$ lives  in the  
algebroid fiber over $X(u)+ \delta_\epsilon X(u)$.   Fortunately, we have  the connection $\nabla^\omega$ which we can use
to parallel transport the vector $A_\epsilon(u)$ from $X(u)+ \delta_\epsilon X(u)$ to $X(u)$; the result of this parallel 
transport we denote as $A_\epsilon ^{\vert\vert}(u)$ and it now makes  perfect sense to define a ``parallel" invariant variation 
$\delta ^{\vert\vert}_\epsilon A$ by the formula
\be \delta ^{\vert\vert}_\epsilon A:=A_\epsilon ^{\vert\vert}(u)-A(u).\label{bbl}\ee
It is clear that  the knowledge of the parallel variation $\delta ^{\vert\vert}_\epsilon A$  fully determines the infinitesimally  
transformed gauge field $A_\epsilon(u)$ and vice versa, since those two quantities are tied  by the parallel transport.  
The concrete  formula for $\delta ^{\vert\vert}_\epsilon A$ is then given by the following nice invariant expression
\be \delta ^{\vert\vert}_\epsilon A=(X^*\nabla^\omega)\epsilon -\ ^Q{\mathcal T}^\omega(A,\epsilon). \label{igt2}\ee
Here $X^*\nabla^\omega$  stands for  the standard pull-back of the connection $\nabla^\omega$   and  the torsion term at 
some $u\in\Sigma$ should be understood as $X^*\ ^Q{\mathcal T}^\omega(\check{A},\check{\epsilon})\big\vert_{X(u)}$, where    
$\check{A}$ and $\check{\epsilon}$   are any sections of the respective bundles $\Omega^1(M,Q)$ and $Q$  fulfilling the properties  
$X^*\check{A} \big\vert_{X(u)}=A(u)$ and $\check{\epsilon} \big\vert_{X(u)}=\epsilon(u)$. As before, 
it  is  the   $C^\infty(M)$-bilinearity  of the torsion which guarantees that the ambiguities in the choices of the lifted sections 
$\check{A}$ and $\check{\epsilon}$ do not influence the value of the parallel variation  $\delta ^{\vert\vert}_\epsilon A$.

The same philosophy we use for the description of the infinitesimal gauge transformation of the Lagrange multiplier $\eta$ 
with the result
\be  \delta ^{\vert\vert}_\epsilon \eta = -\ ^Q{\mathcal T}^{*\omega}_\epsilon\eta.\label{igt3}\ee
Here the operator $\ ^Q{\mathcal T}^{*\omega}_\epsilon: Q^*\to Q^*$  is obtained by the transposition of the $C^\infty(M)$-linear  
operator $\ ^Q{\mathcal T}^{\omega}_\epsilon: Q\to Q$ defined itself  in terms of the torsion as
\be \ ^Q{\mathcal T}^{\omega}_\epsilon s:= \ ^Q{\mathcal T}^{\omega}(\epsilon,s),\quad s\in \Gamma(Q).\ee

Of course, the  invariant formulas can be worked out in components, upon a choice of some local  coordinates 
$X^i$ on $M$ and  local frames $e_a$ on  the algebroid $Q$.\footnote{By the local frames we mean   
$C^\infty(M)$ bases in the spaces of the local sections 
of the algebroid bundle $Q$.}  Explicitly, we introduce $1$-forms $A^a(u)$ on $\Sigma$, 
scalars $\eta_a(u)$ on $\Sigma$, vector fields $v_a$ on $M$,  $1$-forms $\omega^b{}_a$  on $M$ and structure functions
 $C^a{}_{bc}(X)$ on $M$ by the relations:  
\be A=A^a(u)e_a\big\vert_{X(u)},\quad \eta =\eta_a(u)e^*_a\big\vert_{X(u)},  \quad   v_a =\rho(e_a)\,, 
\quad \nabla^\omega e_a=\omega^b{}_a\otimes e_b,  \quad [e_b,e_c]_Q=C^a{}_{bc}(X)e_a\,. \label{fram}\ee
Inserting all those data in our invariant action (\ref{5.1}), we recover straightforwardly  the  component action (\ref{2.1}) of CDJ.

We now give a more detailed calculation, illustrating how to recover
 the   component gauge transformation (\ref{gt2}) from the invariant formulas (\ref{igt1}), 	(\ref{igt2}) and (\ref{igt3}). 
 First we concentrate on the most involved case of the transformation of the gauge field $A$.
 Using the  parametrisation   $A=A^a(u)e_a\big\vert_{X(u)}$, we set, respectively,
\be A_\epsilon(u) \equiv A^a_\epsilon(u)e_a\bigl\vert_{X(u)+\rho(\epsilon)(u)},\quad 
   \delta ^{\vert\vert}_\epsilon A(u)\equiv \delta^{\vert\vert}_\epsilon A^a(u) e_a\bigl\vert_{X(u) }\,, 
   \quad \delta _\epsilon A^a(u):=A^a_\epsilon(u)-A^a(u).\ee
 If the frames $e_a$ were covariantly  constant, then $A_\epsilon ^{\vert\vert}(u)$  would be equal to $A^a_\epsilon(u)e_a\bigl\vert_{X(u) }$, 
 hence, following (\ref{bbl}), the components of the parallel variation $\delta^{\vert\vert}_\epsilon A^a(u)$ would be equal
to the ordinary variations $ \delta _\epsilon A^a(u)$. However, if $e_a$ are not covariantly constant, there is a correction 
proportional to the covariant derivative $\nabla^\omega_\epsilon e_a=\iota_\epsilon\omega^b{}_a e_b$ which we find to be equal to
\be \delta^{\vert\vert}_\epsilon A^a=  \delta _\epsilon A^a+ \iota_\epsilon\omega^a{}_b A^b.  \label{dod}\ee
Now the formula (\ref{igt2})  worked out in components yields
\be  \delta^{\vert\vert}_\epsilon A^a=d\epsilon^a+\omega^a{}_{bi}dX^i \epsilon^b       -\left(\iota_{v_b}\omega^a{}_{c}  -  
\iota_{v_c}\omega^a{}_b-C^a{}_{bc}(X) \right) A^b\epsilon^c.\label{vko}   \ee
Combining the equations (\ref{dod}) and (\ref{vko}), we find
\be \delta _\epsilon A^a =d\epsilon^a+\omega^a{}_{ci}(dX^i-v^i_bA^b)\epsilon^c+C^a{}_{bc}(X)A^b\epsilon^c,\ee
which is nothing but the CDJ gauge transformation (\ref{gt2}).

The case of the field $\eta$ is even simpler. First we find
\be  \delta ^{\vert\vert}_\epsilon \eta_a=  \delta_\epsilon \eta_a-\iota_\epsilon\omega^c{}_a\eta_c,\label{dod'}\ee
and then Eqn.~(\ref{igt3}) written in components gives
\be 
\delta ^{\vert\vert}_\epsilon \eta_a=\left(-C^c{}_{ab}+\iota_{v_a}\omega^c{}_b-\iota_{v_b}\omega^c{}_a\right)\epsilon^b\eta_c\,.\label{vko'}
\ee
Combining the equations (\ref{dod'}) and (\ref{vko'}), we finally  find
\be 
\delta _\epsilon \eta_a = \left(-C^c{}_{ab} \eta_c+v^i_a\omega^c{}_{bi}\eta_c\right) \epsilon^b\,,
\ee
which is nothing but the third of the CDJ gauge transformation (\ref{gt2}).

Next we have to clarify the question of the gauge symmetry of the invariant CDJ  action (\ref{5.1}) with respect 
to the gauge transformations (\ref{igt1}), 	(\ref{igt2}) and (\ref{igt3}). We have already learned from the component 
calculations, that this  gauge symmetry  is not automatic but requires some compatibility of the background data 
$G$,$B$, $Q$ and $\nabla^\omega$.  We have found three compatibility conditions which, written invariantly, become
  
\medskip
\noindent  \begin{enumerate}
  \item For  every section $\chi$ of the  bundle $Q$, it must hold
 \begin{align}\L_{\rho(\chi)}G  & =\left(\rho(\nabla^\omega\chi)\otimes{\rm Id}+{\rm Id}\otimes \rho(\nabla^\omega\chi)\right)G\,,\label{ii1} \\
\L_{\rho(\chi)}B & =\left(\rho(\nabla^\omega\chi)\otimes{\rm Id}+{\rm Id}\otimes \rho(\nabla^\omega\chi)\right)B\,,\label{ii2}
\end{align}
where  $\rho(\nabla^\omega\chi)\in T^*M\otimes TM$ is viewed 
as the linear operator on $T^*M$;

\item The   connection $\nabla^\omega$ on the algebroid bundle  $Q$ must be flat; 
\item For any two sections $s,t\in\Gamma(Q)$, the following implication must hold
\be 
\nabla^\omega s=\nabla^\omega t=0 \implies \nabla^\omega[s,t]_Q=0\,.
\ee
\end{enumerate}
 
If  the conditions (2) and (3) are fulfilled we say that  the  Lie algebroid $(Q,\nabla^\omega)$  is  admissible.
By the component calculations,  we have established that to every admissible   Lie algebroid $(Q,\nabla^\omega)$  there is a naturally associated   Lie algebra $\mathfrak g(Q,\omega)$, 
consisting  of covariantly constant local sections of $Q$. The structure of the Lie algebra $\mathfrak g(Q,\omega)$  is induced from
the Lie algebroid bracket $[\cdot\, ,\cdot]_Q$ and the condition (1) implies that  the 
action of $\mathfrak g(Q,\omega)$  on the sigma model background $(G,B)$---via the anchor map---is isometric.  
Therefore every CDJ theory  is  locally equivalent to the standard 
intermediate gauge theory  used to derive the traditional non-Abelian T-duality.  From the global point of view, 
it may happen that the Lie algebra $\mathfrak g(Q,\omega)$ action on non-simply connected targets $M$ cannot be 
made global by   parallel transport.\footnote{This could happen, for example, by considering  targets   of the form of  
coset of Lie group by its discrete subgroup.} In such a case, the CDJ proposal would give a new insight 
on subtle  topological issues related to the standard non-Abelian T-duality, rather than a recipe to produce
new genuinely non-isometric T-dual pairs of sigma models.  
 
Let us finish this section by stressing that, from the Lie algebroid vantage point,  it does not have  any invariant meaning to say  that
``a non-isometric action of a Lie algebra   on $M$ is gauged".  This is because the local
imput data $M,G,B,v_a,\omega^a{}_{b},C^c{}_{ab}(X)$  used by CDJ to construct their intermediate gauge theory can be equally well replaced 
by  equivalent data $M,G,B,\hat v_a,\hat\omega^a{}_{b},\widehat C^c{}_{ab}(X)$ without changing the T-duality pattern. Given two local anchored frames $v_a$ 
and $\hat v_a$, the structure  {\it constants} $C^c{}_{ab}$ and $\widehat C^c{}_{ab}$ may be related by a {\it non-constant} matrix $K^a{}_b(X)$, 
and as we shall illustrate with examples in the next section,  those two sets of 
structure constants may even define two non-isometrically acting non-isomorphic Lie algebras, the gauging of which  yields the same
T-dual pair of sigma models!    
 
We conclude this section by emphasising that it is solely the Lie algebroid structure $(Q,\nabla^\omega)$ that has  invariant meaning in the CDJ theory. 
Recall, however, that the   gauge invariance of the CDJ theory based on the admissible  Lie algebroid  requires the existence
of the preferred covariantly constant framing on $Q$  for which the anchored  action on the sigma model background $(G,B)$ is isometric.  
Moreover, the structure functions of the covariantly constant frames must be constant
and they thus define the preferred Lie algebra $\mathfrak g(Q,\omega)$. In this case saying   ``the isometric action of the  
Lie algebra   $\mathfrak g(Q,\omega)$ on $M$ is gauged" does have an invariant meaning, 
and it is in this way that the standard non-Abelian T-duality input (i.e. the isometric action  of some Lie algebra on the target) is 
recovered from the structure of the admissible Lie algebroid $(Q,\nabla^\omega)$.
 

\section{Examples} 

Let $M$ be the manifold $\R^3$ parametrised by global Cartesian coordinates $(X^1,X^2,X^3)$, equipped with the metric  
\begin{align}
ds^2 = (dX^1)^2 + \left( dX^2 - X^1 dX^3 \right)^2 + (dX^3)^2\,, \label{metr}
\end{align}
and vanishing $B$-field.

Let  $Q=TM$ be the tangent  Lie algebroid of $M$ (with the identity anchor map) and let $\widehat \nabla$ be a flat connection on $TM$ 
defined by declaring that the following global frame $\hat e_a$  is covariantly constant:
\be \hat e_a= \{\pr_{1}+X^3\pr_{2} , \pr_{2} , \pr_{3}\}\,.
\ee
This structure $(TM,\widehat\nabla)$ defines the admissible Lie algebroid in the sense of Section 5, since the structure 
functions of the covariantly constant framing $\hat e_a$ are all constant. In fact
 the only non-vanishing ones are $\widehat C^2{}_{31}=-\widehat C^2{}_{13}=1$. Let us now establish
that  the Lie derivatives ${\mathcal L}_{\hat e_a}$   of all three differentials 
$dX^1, dX^2 - X^1 dX^3, dX^3$  vanish for every $a=1,2,3$. This can be seen either by a direct calculation or by remarking that the forms
$dX^1, dX^2 - X^1 dX^3, dX^3$ form a basis in the space of the left-invariant  $1$-forms on the
Heisenberg group $\HH$ consisting  of the matrices of the following form:
\begin{align*}
{\HH} =  \left\{
\left(\begin{matrix}
1 & X^1& X^2 \\
0 & 1 & X^3 \\
0 & 0 & 1
\end{matrix}\right), \quad X^1,X^2,X^3\in \R\right \} \,.
\end{align*}
while $\hat e_a$ form the basis in the space of right-invariant vector fields in $\HH$. Either way, we conclude that the Lie derivatives
$\L_{\hat e_a}$  of the metric (\ref{metr}) vanish since the metric is constructed from the left-invariant forms
$dX^1, dX^2 - X^1 dX^3, dX^3$.

We have now all ingredients to define the  Lie algebroid gauged sigma model and its action reads
$$ \hat S(X, \widehat A,\hat\eta)=\jp \int_\Sigma \left( (dX^1 -\widehat A^1)\wedge*(dX^1 -\widehat A^1)+   
(dX^3 -\widehat A^3)\wedge*(dX^3 -\widehat A^3)\right)$$$$ +\jp \int_\Sigma (dX^2-X^3\widehat A^1-
\widehat A^2-X^1(dX^3 -\widehat A^3))\wedge *(dX^2-X^3\widehat A^1-\widehat A^2-X^1(dX^3 -\widehat A^3))$$\be 
+\int_\Sigma (\hat \eta_1 d\widehat A^1 +\hat\eta_2(d\widehat A^2+\widehat A^3\wedge\widehat A^1)
 +\hat \eta_3 d\widehat A^3)\,.\label{hrc}\ee
The gauge transformations are
\be \delta_\epsilon X^1=\epsilon^1, \quad  \delta_\epsilon X^2=\epsilon^2+X^3\epsilon^1, \quad \delta_\epsilon X^3=\epsilon^3, \nonumber\ee
 \be \delta_\epsilon \widehat A^1=d\epsilon^1, \quad  \delta_\epsilon \widehat A^2=d\epsilon^2+\widehat A^3\epsilon^1
 -\widehat A^1\epsilon^3, \quad \delta_\epsilon\widehat A^3=d\epsilon^3, \label{pog2}\ee
  \be \delta_\epsilon \hat \eta_1=\epsilon^3\hat\eta_2, \quad  \delta_\epsilon \hat \eta_2=0,\quad
 \delta_\epsilon \hat \eta_3=-\epsilon^1\hat\eta_2\,. \nonumber\ee
Since we are gauging the isometry, we are doing the standard non-Abelian T-duality, nevertheless, for the sake of illustration, 
we go on further with the well-known procedure how to recover from (\ref{hrc}) the T-dual pair of sigma models.

Varying the Lagrange multipliers $\hat\eta_a$ we obtain the Maurer Cartan relations
\be d\widehat A^1=0\,, \quad d\widehat A^2+\widehat A^3\wedge\widehat  A^1=0\,,\quad d\widehat A^3=0 \,.
\label{mac}\ee
 which can be solved in full generality as
 \be \widehat A^1=-dx^1,\quad \widehat A^2=-dx^2+x^1dx^3,\quad  \widehat A^3=-dx^3.\label{smc}\ee
Plugging the solution (\ref{smc})  into the action (\ref{hrc}), we recover the original sigma model
corresponding to the metric (\ref{metr}) and the vanishing B-field:
\be \widehat S(Y^k)=\jp \int_\Sigma \left( dY^1 \wedge *dY^1+ (dY^2- Y^1dY^3)\wedge *(dY^2- Y^1dY^3)+  dY^3 \wedge *dY^3     \right),\label{orig}\ee
where\footnote{The field redefinitions (\ref{frd}) reflect the group multiplication law in the Heisenberg
group $\HH$ and the new fields $Y$ can be interpreted as $X$ acted upon by the non-infinitesimal gauge transformation
``$x$'' obtained by exponentiating the gauge transformations  (\ref{pog2}).}
\be Y^1:= X^1+ x^1\,,\quad Y^2 := X^2+x^2 + x^1X^3\,,\quad Y^3:=X^3+x^3\,.\label{frd}\ee

In order to recover the dual sigma model, we first make in (\ref{hrc}) the following field redefinitions:
\be \widehat B^1:=\widehat A^1-dX^1,\quad \widehat B^2:=\widehat A^2-dX^2+X^3\widehat A^1+X^1(dX^3-\widehat A^3),\quad 
\widehat B^3:=\widehat A^3-dX^3,\ee
\be \hat\mu_1:=\hat\eta_1-X^3\hat\eta_2,\quad \hat\mu_2:=\hat\eta_2,\quad \hat\mu_3:=\hat\eta_3+X^1\hat\eta_2,\ee
in terms of which the action  (\ref{hrc}) becomes
\be 
\widehat S(\widehat B,\hat\mu)=\jp \int_\Sigma (\widehat B^1\wedge* \widehat B^1+\widehat B^2\wedge* \widehat B^2 +\widehat B^3\wedge* \widehat B^3)-\int_\Sigma (
d \hat \mu_1 \wedge \widehat B^1 +d\hat\mu_2\wedge \widehat B^2    +d \hat \mu_3 \wedge \widehat B^3-\hat\mu_2 \widehat B^3\wedge\widehat  B^1)
\,.\label{hrc'} \ee
Varying $\widehat B$ in $\widehat S( \widehat B,\hat\mu)$ gives
\be 
\widehat B^1=-\frac{1}{1+\hat\mu^2_2}(\hat\mu_2d\hat\mu_3+*d\hat\mu_1), \quad \widehat B^2=
-*d\hat\mu_2,\quad \widehat B^3=\frac{1}{1+\hat\mu^2_2}(\hat\mu_2d\hat\mu_1-*d\hat\mu_3).\label{bsl}  \ee
Inserting (\ref{bsl}) back into $\widehat S(  \widehat B,\hat\mu)$ gives the dual sigma model:
\be 
\widehat S(\hat\mu)=\jp \int_\Sigma \frac{1}{1+\hat\mu^2_2}\left(d\hat\mu_1\wedge *d\hat\mu_1
+(1+\hat\mu^2_2)d\hat\mu_2\wedge *d\hat\mu_2+d\hat\mu_3\wedge *d\hat\mu_3+2\hat\mu_2d\hat\mu_1\wedge d\hat\mu_3\right) \,.\label{td}
\ee
 
There is a simple reason why we have chosen to work out this particular example of the non-Abelian T-duality. It is because CDJ  have applied   in \cite{CDJ15}   
their ``T-duality without isometry" recipe exactly on the  sigma model background  (\ref{metr}).\footnote{CDJ wrote in \cite{CDJ15} that they dualised the    
Heisenberg nilmanifold but the computation that they performed therein
concerns, in reality,   the  Heisenberg group target.}   Our point is to emphasize that 
CDJ  did not obtain anything new but just
the same dual model (\ref{td})  as we did by gauging the   isometry(cf.\ Eqn.\ (4.19) of Ref.\ \cite{CDJ15}).
We also understand the  reason why this fact is not coincidental. Indeed,  it occurs  because CDJ   just used a different framing in order 
to write the same  invariant Lie algebroid $(TM,\hat\nabla)$  gauge theory in components.   Actually, the  framing  of CDJ was formed by the following basis 
of the left-invariant vector fields on the group $\HH$
\be e'_a=\{\pr_{1}, \pr_{2}, X^{1}\pr_{2}+\pr_{3}\}, \ee
and because the Lie derivatives of the left-invariant forms with respect to the left-invariant vector fields do not vanish (for   
non-Abelian Lie groups),  the background sigma model metric (\ref{metr}) is not invariant.  Of course, the non-invariance of the metric can be   seen also 
from the formula (\ref{ii1}) since the framing $e'_a$ is not covariantly constant.  

CDJ used the frame $e'_a$ to carry out the non-isometric
gauging  of the Heisenberg group acting on itself from the right, but we have  already explained in Section 5 that only gauging with respect  
to the covariantly constant frame has invariant meaning. In fact, the gauging of CDJ is not even the only possible non-isometric gauging 
based on an action of a Lie group.  Indeed, we can even consider the   non-isometric gauging of the sigma model
(\ref{orig}) 
by the action of the Abelian translation group $\R^3$ on itself! We achieve that  by choosing yet a third frame $e_a$ for which the 
structure functions  $C^a{}_{ bc}$ all vanish. It reads simply 
\be e_a=\{\pr_{1}, \pr_{2},  \pr_{3}\}.\quad \ee
The framing $e_a$ is neither covariantly constant nor does it leave the metric (\ref{metr}) invariant.  
{}From the relation
\be \hat e_a=e_b(K^{-1})^b{}_a, \quad K= \left(\begin{matrix}
1 & 0& 0 \\
-X^3 & 1 & 0 \\
0 & 0 & 1
\end{matrix}\right),\ee
it easily follows that the input data of the CDJ gauge theory is the metric (\ref{metr}), $B=0$, the frame
$e_a$ and the matrix valued $1$-form $\omega$ given by
\be \omega= \left(\begin{matrix}
0 & 0& 0 \\
-dX^3 & 0 & 0 \\
0 & 0 & 0
\end{matrix}\right).\ee
The action (\ref{2.1}) of the CDJ gauge theory can be specified for those 
data  and the direct calculation  yields, without surprise,  again the dual pair (\ref{orig}), (\ref{td}) of sigma models.

\vskip1pc

\noindent {\bf Remark}. We note that whenever an $n$-dimensional group manifold $\mathsf G$ admits a global 
coordinate system $X^1,...,X^n$, then every $\mathsf G$-isometric background can  be either   T-dualised
by the standard isometric gauging based on the $\mathsf G$ action on itself or by 
the CDJ non-isometric gauging of the Abelian group $\R^n$ generated by the coordinate vector
fields $\pr_1,...,\pr_n$. 

\vskip1pc

Our second example directly generalizes the first one in the sense we replace the Heisenberg group $\HH$
by an arbitrary Lie group  $\mathsf G$.  We shall work invariantly since we shall no longer need
to make comparison  with the coordinate calculations of CDJ.

Our Lie algebroid $Q$ is now  the tangent bundle $T\mathsf G$ and we pick the flat connection $\nabla^\omega$ by declaring that 
the right-invariant vector fields on  $\mathsf G$  are covariantly constant.  We write the sigma model action, which we want to gauge, as 
\be S(g) =\jp\int_\Sigma(g^{-1}dg\wedge *g^{-1}dg)_G+\jp\int_\Sigma(g^{-1}dg\wedge g^{-1}dg)_B.\label{exs}\ee
Here $g^{-1}dg$ is the left-invariant Maurer-Cartan form on $\GG$  and the background geometry is encoded  in the choice 
of two non-invariant bilinear forms $(\cdot\, ,\cdot)_G$ and $(\cdot \, , \cdot)_B$ defined on the Lie algebra $\g$ of a Lie group $\GG$, 
the former   symmetric and the latter antisymmetric.  
In other words,  the metric $G$ and the $2$-form $B$ underlying this particular sigma model action  are obtained
by the left-transport of the bilinear forms $(\cdot \, , \cdot)_G$ and $(\cdot \, , \cdot)_B$ to every point of the group manifold.

The CDJ gauge theory corresponding to the right-invariant framing is now the standard Ro\v cek-Verlinde  
Yang-Mills theory, traditionally  used to work out the standard non-Abelian T-dual of the model (\ref{exs}). It can be also obtained without knowing
anything about Lie algebroids, just by gauging the rigid left action of the group $\GG$ on itself which  is the symmetry of the action (\ref{exs})  :
\be S(g,\widehat A,\eta)=\jp\int_\Sigma(g^{-1}Dg\wedge *g^{-1}Dg)_G+\jp\int_\Sigma(g^{-1}Dg\wedge g^{-1}Dg)_B
+\int_\Sigma \langle \hat\eta, d\widehat A-\widehat A\wedge \widehat A\rangle.\label{ga}\ee
Here $\widehat A$ and $\hat\eta$ are respectively $\g$-valued $1$-form and $\g^*$-valued $0$-form on the world-sheet 
$\Sigma$ and the covariant derivative is defined as
\be g^{-1}Dg:= g^{-1}dg-g^{-1}\widehat Ag.\ee
Note also the form of  the field strength $d\widehat A-\widehat A\wedge \widehat A$.  
{}From  the  (invariant)  CDJ point of view, it comes from the fact that the torsion of the flat connection leaving 
the right-invariant vector fields covariantly constant coincides precisely with the
commutator on the Lie algebra $\g$. For that matter, this  is coherent with the appearance of the minus sign in the field strength 
which reflects the fact that the structure constants corresponding to the infinitesimal left action  of $\GG$ on itself pick a minus sign with respect
to the structure constants of the Lie algebra $\g$.

The gauged action (\ref{ga}) has the following left gauge symmetry
\be 
(g,\widehat A,\hat\eta)\to (\hat hg,   \hat h\widehat A\hat h^{-1} +d\hat h\hat h^{-1}, {\rm Ad}^*_{\hat h}\hat\eta)\,, \label{gsn}
\ee
or, infinitesimally,
\be \delta_{\hat\epsilon}(g,\widehat A,\hat\eta)=(\hat\epsilon g,d\hat\epsilon -[\widehat A,\hat\epsilon], {\rm ad}^*_{\hat\epsilon}\hat\eta).\ee
Here $\hat h$ and $\hat\epsilon$ are smooth maps from the world-sheet $\Sigma$ to the Lie group $\GG$
and the Lie algebra $\g$, respectively. Note in particular, that the expression $g^{-1}Dg$ turns out to be gauge invariant, which immediately explains the
gauge invariance of the part of the action (\ref{ga}) not containing the Lagrange multiplier.

We now switch from  the natural
isometric  right-invariant framing  to  non-natural non-isometric left-invariant one, and  wish to rewrite the CDJ gauge theory (\ref{ga}) accordingly.  
To work this out,  we can  depart directly  from the invariant action (\ref{5.1}), but it is simpler to do it by  making the 
appropriate field redefinitions  in  the  action  (\ref{ga}), induced by the change of the frame.  For that, it is enough to note that  the  
``new" gauge transformation now  hits the sigma model configuration $g$ from the
right, which implies that the $g$-dependent  frame-changing operator $K$ in Eqn.\ (\ref{mar}) is simply  $K={\rm Ad_{g}}$.
Indeed, writing 
\be  
h= g^{-1}\hat h g\,,
\ee
gives infinitesimally
\be  
\epsilon={\rm Ad_{g^{-1}}} \hat \epsilon , 
\ee
which, following Eqn.~(\ref{mar}), fixes $K={\rm Ad_{g}}$.

With this choice of $K$ the remaining field redefinitions are dictated by Eqn.\ (\ref{fir}):
\be  
A = g^{-1}\widehat Ag, \quad \eta={\rm Ad}^*_{g^{-1}}\hat\eta \,.
\ee
The covariant derivative $g^{-1}Dg$, the action  (\ref{ga}) and the gauge tranformations (\ref{gsn}) then become, respectively,
\be 
g^{-1}Dg= g^{-1}\,dg - A \,,
\ee
\begin{align}
S(g,A,\eta)  = & \jp\int_\Sigma(g^{-1}Dg\wedge *g^{-1}Dg)_G   +\jp\int_\Sigma(g^{-1}Dg\wedge g^{-1}Dg)_B \nonumber \\
& +\int_\Sigma \langle \eta, dA+A\wedge A+g^{-1}Dg\wedge A+A\wedge g^{-1}Dg\rangle\,.\label{ga'}
\end{align}

\be 
(g, A, \eta)\to ( gh, A-g^{-1}dg +(gh)^{-1}\, d(gh), \eta)\,.  \label{gsn'}
\ee
Note that the infinitesimal version of the gauge transformations (\ref{gsn'})  read
\be 
\delta_\epsilon(g, A, \eta) =( g\epsilon, d\epsilon +[A,\epsilon] +{\rm Ad}_{g^{-1}Dg}\epsilon, 0 )\,.\label{igsn'}
\ee
We remark that the gauged sigma model action (\ref{ga'}), as well as the infinitesimal 
gauge transformations (\ref{igsn'}), have now indeed the CDJ form (\ref{2.1}), (\ref{gt2}),
where the $1$-form $\omega$ with values in End$(\g)$ is invariantly written as
\be 
\omega={\rm ad}_{g^{-1}dg}\,.
\ee
If a reader would look just  at the gauge theory action (\ref{ga'}) as well as at the infinitesimal gauge transformations 
(\ref{igsn'}),  without knowing how we have obtained them, they would probably believe
they have some exotic gauging of the right action of the group $\GG$ on itself. 
Our point is that this CDJ  non-isometric exotic   right gauging is just the standard isometric left gauging in disguise, therefore
it cannot give rise to any new T-duality pattern.
 

\section{Conclusion and outlook}

We have ruled out the proposal of Chatzistavrakidis, Deser and Jonke in the sense that it
does not give rise to a new pattern of genuinely non-isometric T-duality.  Nevertheless, we see some room to apply the invariantly 
formulated CDJ gauge theory  of Section 5 to study
some subtle topological effects within the framework of the traditional non-Abelian T-duality.
This possibility might take place if there exist   admissible Lie algebroids $(Q,\nabla^\omega)$  for which the
invariant Lie algebra $\g(Q,\omega)$  would act just locally on the target manifold $M$ and could  not 
be extended to a global action. It is plausible to expect that this situation may occur for which the targets $M$ are of the form of the quotient of a
Lie group by one of its discrete subgroups. The CDJ theory  could then take into account the phenomena
of winding strings on non-contractible cycles of $M$.


 \section*{Acknowledgements}
PB is supported by the Australian Government through the Australian Research Council's Discovery 
Projects funding scheme (projects DP150100008 and DP160101520).
CK  acknowledges the financial support provided by the LIA scientific 
cooperation program of Australian National University in 
Canberra and CNRS in France.


\end{document}